# Synthesis of anti-perovskite-type carbides and nitrides from metal oxides and melamine

Daigorou Hirai,* Hidetake Tanaka and Zenji Hiroi

*Institute for Solid State Physics, University of Tokyo, Kashiwa, Chiba 277-8581, Japan*
*e-mail address: dhirai@issp.u-tokyo.ac.jp*

Four anti-perovskite-type compounds, $ZnNNi_3$, $ZnCNi_3$, $SnNCo_3$, and $SnCCo_3$, are synthesised through reactions between ingredient metal oxides and organic compound melamine ($C_3H_6N_6$). $ZnNNi_3$ and $ZnCNi_3$ are selectively synthesised by choosing different reaction temperatures and nominal oxide–to–melamine ratios. $SnNCo_3$ is synthesised for the first time by this melamine method. Resistivity, magnetisation, and heat capacity measurements reveal that $SnNCo_3$ is a correlated metal with a high density of states at the Fermi level. Our results demonstrate that this feasible synthetic route using melamine is useful in the search for complex metal carbides and nitrides toward novel functional materials.

## I. INTRODUCTION

Anti-perovskite-type compounds with a general formula of $AXM_3$ (A = Ga, Al, In, Zn, Sn, etc.; X = C, N, B; M = transition metal) have been intensively studied owing to their unique physical properties and functionalities such as superconductivity[1–3], negative thermal expansion[4–7], giant magnetoresistance[8–10], magnetostriction effect[11,12], magnetocaloric effect[13–15], thermoelectric property[16,17], and temperature coefficient of resistivity of approximately zero[18]. As illustrated in Fig. 1(a), the crystal structure is closely related to the perovskite structure with the formula of $AMX_3$. In the perovskite structure, the A and M sites are occupied by cations, where the M cation is octahedrally coordinated by six anions at the X sites, while in the anti-perovskite structure, an anion occupying the X site is surrounded by six M cations to form an $XM_6$ octahedron. Many anti-perovskite-type compounds have been synthesised; however, considering the flexibility of the anti-perovskite-type structure, we expect that there are more compounds of this type that have not been yet studied.

Most of the anti-perovskite-type nitrides were synthesised by reaction of metal oxides with ammonia ($NH_3$) gas at high temperatures, typically around 800 °C, which is usually referred to as ammonolysis[19]. The reaction is driven by the high nitriding activity of ammonia; the high temperature is required to accelerate the solid-state reaction with the oxides. However, as ammonia is decomposed into chemically inactive nitrogen and hydrogen gases above 300 °C[19], the actual ammonolysis is performed under a non-equilibrium condition in a bulk flow of ammonia gas in its decomposition process. Such a non-equilibrium reaction is difficult to control, and it is not often simple to obtain the desired nitride in a single phase.

Recently, new synthetic routes through reactions between organic reagents and metal oxides have been intensively investigated to overcome the difficulties of the non-equilibrium gas–solid reactions[20–24]. For the synthesis of nitrides, amide ($NaNH_2$)[25], azide ($NaN_3$)[26], and urea ($CO(NH_2)_2$)[27–30] are used as the nitriding reagent. As the reaction is performed in an enclosed tube, the ratio of reactant and nitriding reagent can be precisely controlled. Therefore, the reaction is highly reproducible compared to ammonolysis.

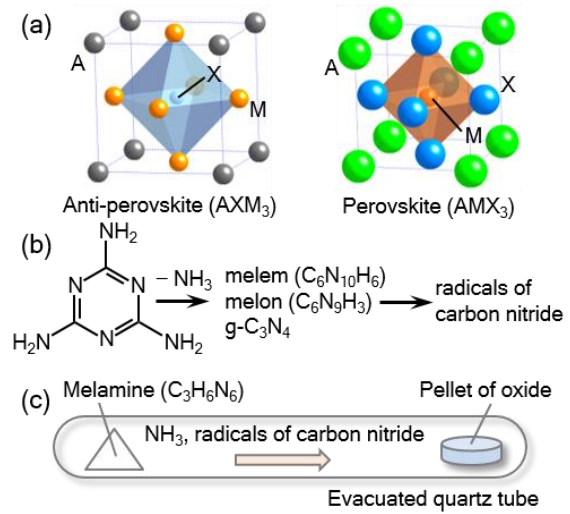

Fig. 1 (a) Crystal structures of the anti-perovskite-type compound $AXM_3$ and related perovskite $AMX_3$. (b) Melamine ($C_3H_6N_6$) gradually decomposes upon heating, releasing the reactive $NH_3$ gas, which is decomposed into radicals of carbon nitrides. (c) Reaction tube. $NH_3$ and radicals of carbon nitride from melamine react with oxides to generate an anti-perovskite-type nitride or carbide.

In this study, we focus on reactions using melamine ($C_3H_6N_6$)[31], which is a widely used raw material, chemically stable in air at ambient temperature[32]. Binary nitrides of eight elements, Ga, Cr, B, Al, Ti, V, Nb, and Ta, were successfully prepared by reaction between melamine and oxide precursors[31]. Upon heating, melamine decomposes into g-$C_3N_4$ releasing ammonia gas ($NH_3$) below 560 °C. A further heating leads to decomposition of

g-C$_3$N$_4$ into reactive radical species containing hydrogen or carbon (Fig. 1(b))[33–35]. It is considered that these reactive radicals form thermodynamically stable CO and H$_2$O, which help the nitriding reactions. The above nitrides can be synthesised by the melamine method at significantly lower temperatures compared to that of the conventional ammonolysis. It is worth noting that at higher temperatures and higher nominal melamine contents, binary carbides of Mo, W, V, Nb, and Ta were synthesised by the melamine method.[36] This promising synthetic route may be applied to more complex compounds, which could reveal new compounds with novel functionalities.

In this study, we report the synthesis of anti-perovskite-type nitrides and carbides through reactions between the ingredient metal oxides and melamine, and their physical properties. ZnNNi$_3$ and ZnCNi$_3$ are selectively synthesised at different reaction temperatures and nominal melamine contents. A new anti-perovskite-type compound SnNCo$_3$ is successfully synthesised, which is a correlated metal with a Curie–Weiss-type magnetic susceptibility and large electronic specific heat coefficient, revealed by heat capacity measurements. Our findings demonstrate that the melamine method is applicable to not only simple binary systems but also complex systems. The reaction using melamine is valuable for further studies on novel functional nitrides and carbides.

## II. EXPERIMENTAL

In a typical synthesis, powders of ingredient oxides were mixed in an agate mortar and pressed into a pellet. The pellet was placed at one side of an evacuated quartz tube (typically 70 mm in length, 12 mm in inner diameter, and 14 mm in outer diameter) with melamine at the other side, as shown in Fig. 1(c). The relative amount of melamine was varied as listed in Table I. The quartz tube was heated at reaction temperature for 12 hours. The total amount of melamine was adjusted so that the pressure in the tube at the temperature was small enough to avoid an explosion.

X-ray powder diffraction (XRD) measurements were conducted in a RINT–2500 diffractometer (Rigaku) using Cu-$K\alpha$ radiation. The data was collected at 298 K over a $2\theta$ range between 10 and 120° with a step size of 0.02°. Lattice parameters and atomic displacement factors were refined by the Rietveld method using the FULLPROF program. The background was characterized by use of a six-coefficient polynomial function. Magnetization measurements were performed in a Magnetic Property Measurement System MPMS3 (Quantum Design). Resistivity and heat capacity measurements were conducted in a Physical Property Measurement System PPMS (Quantum Design).

## III. RESULTS AND DISCUSSION

### 1 Selective syntheses of ZnNNi$_3$ and ZnCNi$_3$

We apply the melamine method for preparation of the known anti-perovskite-type nitride and carbide ZnXNi$_3$ (X = C and N). In the previous reports, melamine and oxides were mixed into a pellet[31,36], while in this study, a pelletised oxide is placed separately from melamine in an evacuated quartz tube so that gas-phase products from the decomposition of melamine react with the oxides. This method is suitable for measurements of physical properties as it reduces the contamination of products by decomposition residuals.

Three fabrications were performed with a fixed stoichiometric metal ratio (ZnO:NiO = 1:3) at different melamine contents and reaction temperatures for 12 h. The product obtained in condition 1 with a ZnO:NiO:C$_3$N$_3$(NH$_2$)$_3$ ratio of 1:3:3 at 600 °C exhibits the XRD pattern shown in Fig. 2(a). All of the diffraction peaks are indexed to a cubic lattice at extinctions consistent with the space group $Pm\bar{3}m$, which indicates that the product is synthesised in a single phase. The lattice constant of the product is $a$ = 3.7648(3) Å, which is comparable to that of the anti-perovskite-type nitride ZnNNi$_3$ ($Pm\bar{3}m$) of $a$ = 3.756 Å[2]. Other candidates are alloys of Ni and Zn, and carbides and nitrides of Ni or Zn. However, they do not exhibit equal or similar XRD patterns to that in the figure. Therefore, we can conclude that ZnNNi$_3$ was successfully

Table 1 Starting composition, reaction temperature, products, and lattice parameters for the anti-perovskites synthesised in this study.

| Starting composition | Temperature (°C) | Products | Lattice parameter | Reported values |
|---|---|---|---|---|
| ZnO + 3NiO + 3C$_3$H$_6$N$_6$ | 600 | ZnNNi$_3$ | 3.7648(3) | 3.756 |
| ZnO + 3NiO + 5C$_3$H$_6$N$_6$ | 650 | ZnCNi$_3$ | 3.6601(7) | 3.66 |
| ZnO + 3NiO + 5C$_3$H$_6$N$_6$ | 600 | ZnNNi$_3$, ZnCNi$_3$ | 3.7559(4) 3.6601(6) | 3.756 3.66 |
| SnO$_2$ + 3CoO + 5C$_3$H$_6$N$_6$ | 1000 | SnCCo$_3$ | 3.8118(3) | 3.8047(3) |
| SnO$_2$ + 3CoO + 5C$_3$H$_6$N$_6$ | 550 | SnNCo$_3$ | 3.8513(1) | – |

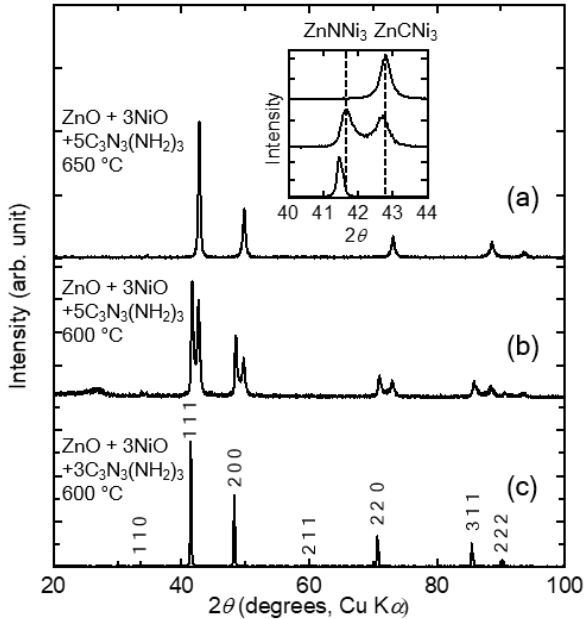

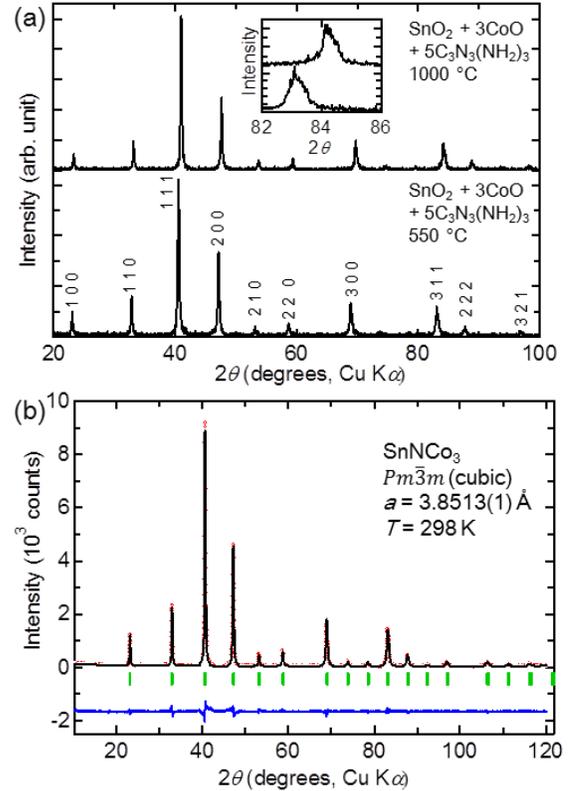

Fig. 2 Powder XRD patterns of the products from ZnO, NiO, and melamine prepared under different conditions. Monophasic samples of ZnNNi$_3$ in condition 1 (a) and ZnCNi$_3$ in condition 3 (c), and their mixtures in condition 2 (b) are obtained. The top inset presents an enlarged view showing differences in peak position between the three samples. The two broken lines represent the peak positions for ZnNNi$_3$ and ZnCNi$_3$

In condition 2, the nominal melamine content was increased, yielding a ratio of ZnO:NiO:C$_3$H$_6$N$_6$ of 1:3:5; the reaction was performed at 600 °C for 12 h. The XRD pattern of the product in Fig. 2(b) is different from that in condition 1. In addition to the peaks from ZnNNi$_3$ obtained in condition 1, a set of peaks with similar relative intensities appear at higher angles. This indicates that two phases with similar cubic structures and different lattice constants exist in the sample. The obtained lattice constants are $a = 3.7559(4)$ Å, which corresponds to ZnNNi$_3$, and $a = 3.6601(6)$ Å.

In condition 3, the nominal melamine content was equal to that in condition 2 (ZnO:NiO:C$_3$H$_6$N$_6$ = 1:3:5), while the reaction temperature was higher (650 °C). The XRD pattern of the product in Fig. 2(c) contains peaks corresponding only to those from the second cubic phase obtained in condition 2 with the smaller lattice constant. Therefore, an almost single cubic phase with the lattice constant of $a = 3.6601(7)$ Å is obtained in condition 3. The lattice constant is equal to $a = 3.66$ Å for the anti-perovskite-type carbide ZnCNi$_3$ ($Pm\bar{3}m$)[37]; no other known products reproduce the XRD pattern. Therefore, ZnCNi$_3$ is synthesised by the melamine method at an excess melamine content and high reaction temperature.

Fig. 3 (a) Powder XRD patterns of SnNCo$_3$ and SnCCo$_3$ prepared under conditions 4 and 5 by the melamine method, respectively. The inset shows an enlarged view of a $2\theta$ range around the (311) diffraction peak. (b) Rietveld refinement for SnNCo$_3$. The red circles, black solid line, blue solid line, and green tick marks represent experimental data, simulation pattern, differences, and diffraction positions, respectively. The atomic positions are: Sn 1$a$ (0 0 0), Co 3$c$ (0 0.5 0.5), and N 1$b$ (0.5 0.5 0.5). The Rietveld agreement factors are: $\chi^2 = 3.33$, $R_{Bragg} = 0.040$, $R_P = 0.086$, and $R_{WP} = 0.126$.

The products and synthetic conditions are summarised in Table 1. Ternary Zn–Ni carbides and nitrides are selectively synthesised by the melamine method. The reaction mechanism will be discussed below. As complete phase separation is observed at the intermediate condition 2, there must be a miscibility gap between them at this synthetic condition. The lattice constants of the two phases in condition 2 are almost equal to those of the pure carbide and nitrides.

### 2 Syntheses of a new nitride SnNCo$_3$ and related carbide SnCCo$_3$

We have applied our melamine method to search for new anti-perovskite-type compounds and successfully obtained the nitride SnNCo$_3$ and related carbide SnCCo$_3$. In a similar reaction protocol, melamine and pelletised mixture of SnO$_2$ and CoO were sealed in an evacuated quartz tube at a molar ratio of $SnO_2:CoO:C_3H_6N_6 = 1:3:5$ and heated at 550 °C for 12 h (condition 4). A single-phase compound crystallised in a cubic symmetry ($Pm\bar{3}m$) is obtained, as shown in the powder XRD pattern in Fig. 3(a). When the reaction temperature was increased to 1000 °C with the same nominal composition (condition 5), a compound with the same cubic symmetry but with a smaller lattice constant was obtained in a single phase. The lattice constants are $a = 3.8513(1)$ Å and $a = 3.81181(3)$ Å for conditions 4 and 5, respectively. The XRD pattern of the product in condition 5 is very similar to that of the reported anti-perovskite-type carbide $SnCCo_3$, while the lattice constant is close to that of the stoichiometric $SnCCo_3$ of $a = 3.8047(3)$ Å[16], larger than that of $SnCCo_3$ with a carbon deficiency of $a = 3.78$ Å[38]. Therefore, $SnCCo_3$ with a minimum number of carbon defects was synthesised in this study.

The similar powder XRD pattern of the product in condition 4 suggests that another anti-perovskite-type compound is synthesised. Considering the lower reaction temperature and larger lattice constant, the product is likely to be $SnNCo_3$. In order to confirm the formation of $SnNCo_3$, the XRD pattern is fitted to a structural model based on the crystal structure of $SnCCo_3$ by the Rietveld method. The fitting is excellent, as shown in Fig. 3(b), even though there are only a small number of refinable parameters with all of the positions of the atoms fixed in the anti-perovskite-type structure. Therefore, we can conclude that a new anti-perovskite-type nitride $SnNCo_3$ was obtained in condition 4.

The lattice constants of $SnXCo_3$ (X = N and C) are larger than those of $ZnXNi_3$ (X = N and C). The metallic radii of the B-site cations Ni (1.149 Å) and Co (1.157 Å) are comparable to each other, while the radius of Sn (1.412 Å) in the A-site is significantly larger than that of Zn (1.249 Å)[39]. Therefore, the difference in lattice constant originates from the difference in size of the A-site ions. On the other hand, a comparison between the nitrides and carbides shows that the lattice constant of $ZnNNi_3$ is 2.86% larger than that of $ZnCNi_3$, while the difference between $SnNCo_3$ and $SnCCo_3$ is 1.04%. These differences originate from the size difference between nitrogen and carbon atoms. In the case of $SnXCo_3$, the effect of size difference seems to be suppressed in the expanded lattice by the large Sn cation.

## 3 Physical properties of the new anti-perovskite-type nitride $SnNCo_3$

The physical properties of $SnNCo_3$ were characterised by resistivity, magnetisation, and heat capacity measurements.

As shown in Fig. 3, our $SnCCo_3$ sample exhibits metallic conductivity with a temperature dependence and magnitude similar to those reported previously[16]. $SnNCo_3$ also exhibits a metallic resistivity, two orders of magnitude larger than that of $SnCCo_3$, which may be attributed to grain boundary scattering as $SnNCo_3$ was poorly sintered at a lower temperature. A small upturn is observed below 25 K in the resistivity of SnNCo$_3$, which may be due to weak localisation by the certain disorder. The observed metallic behaviour should be intrinsic considering that all related anti-perovskite-type compounds including ZnXNi$_3$ (X = N and C)[2,37] are metals. There is no anomaly indicative of a structural or electronic transition such as superconductivity in the resistivity curve of SnXCo$_3$ (X = N and C) above 2 K.

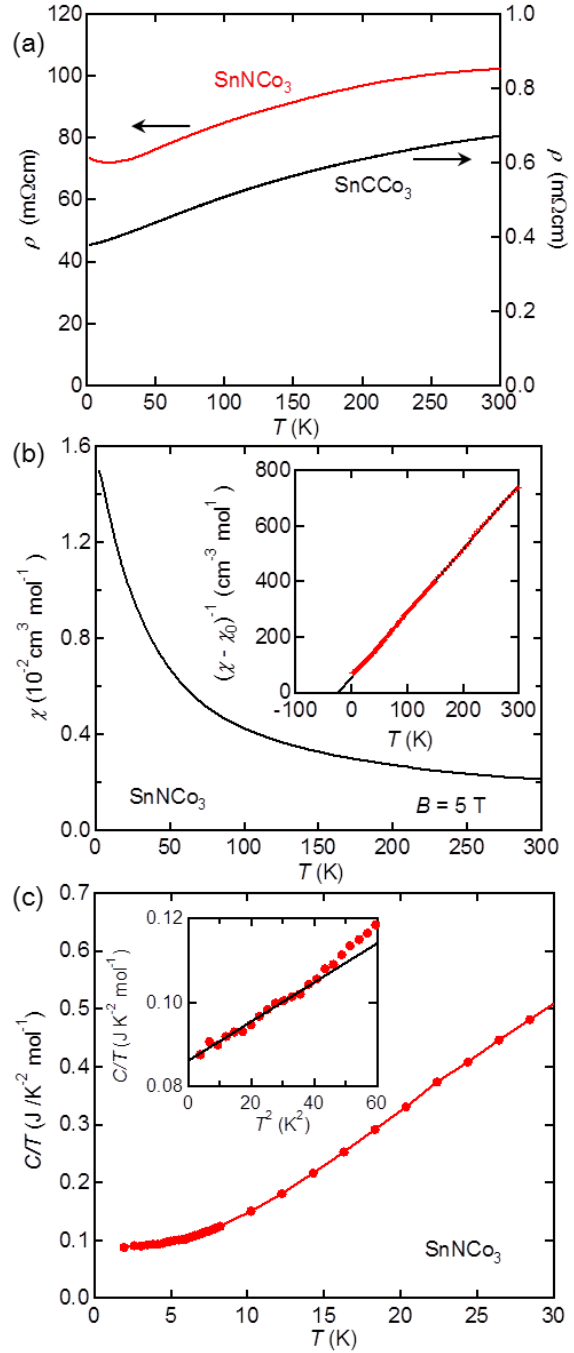

Fig. 4. Fig. 4 Temperature dependences of the (a) resistivity, (b) magnetic susceptibility, and (c) heat capacity of $SnNCo_3$. The resistivity data of $SnCCo_3$ synthesised in this study are also shown in (a). The inset in (b) shows the inverse of the susceptibility after the subtraction of the temperature-independent term (red dots) with a Curie–Weiss fit indicated by the solid black line. The dependence of $C/T$ on $T^2$ is shown in the inset of (c).

SnNCo$_3$ is paramagnetic down to 2 K with no anomaly indicative of a magnetic transition in the temperature dependence of the magnetic susceptibility (Fig. 4(b)). The isothermal magnetic susceptibility at 300 K exhibits a linear dependence on the applied field, which confirms the absence of a ferromagnetic component originating from impurity phases such as elemental Co. As shown in Fig. 4(b), the magnetic susceptibility rapidly increases with the decrease in the temperature, which follows the Curie–Weiss behaviour expected for a localised magnet or correlated metal: $\chi = (T - \Theta_W)/C + \chi_0$, where $C$ is the Curie constant, $\Theta_W$ is the Weiss temperature, and $\chi_0$ is a temperature independent constant. A fitting to the data between 20 and 300 K yields $C = 0.428(2)$ cm$^3$ K mol$^{-1}$, $\Theta_W = -22.3(2)$ K, and $\chi_0 = 7.68(11) \times 10^{-4}$ cm$^3$ mol$^{-1}$. The effective magnetic moment derived from the Curie constant is $\mu_{eff} = 1.85$ $\mu_B$ per formula unit including three Co ions, which is significantly smaller than expected for insulating Co magnets or ferromagnetic Co metal. Therefore, SnNCo$_3$ is a strongly correlated itinerant magnet with dominant antiferromagnetic interactions, where a narrow band attributed to $d$-orbitals with a large density of states (DOS) is located at the Fermi energy ($E_F$).

The high DOS at the $E_F$ in SnNCo$_3$ is demonstrated by heat capacity measurements. As shown in Fig. 4(c), the low-temperature heat capacity data between 2 and 6 K are well fitted to the equation: $C/T = \gamma + \beta T^2$, which yields $\gamma = 86.1(4)$ mJ K$^{-1}$ mol$^{-2}$ and $\beta = 0.465(18)$ mJ K$^{-1}$ mol$^{-4}$. The finite electronic specific heat coefficient $\gamma$ indicates the existence of a Fermi surface in SnNCo$_3$, which is consistent with the metallic resistivity. The $\gamma$ value is significantly higher than those of a conventional metal and related compounds such as SnCCo$_3$ ($\gamma = 40.51(2)$ mJ K$^{-2}$ mol$^{-1}$)[16] and ZnNNi$_3$ ($\gamma = 13$ mJ K$^{-2}$ mol$^{-1}$)[2]. The large $\gamma$ further confirms the correlated metallic state with a high DOS at the $E_F$ realised in SnNCo$_3$. The Debye temperature of SnNCo$_3$ calculated from the coefficient $\beta$ is $\Theta_D = 275(3)$ K, which is close to those of related anti-perovskite-type compounds ZnN$_y$Ni$_3$ ($\Theta_D = 336$ K) and CdCNi$_3$ ($\Theta_D = 352$ K)[2].

According to first-principle calculations for Co-based anti-perovskite-type nitrides[40], the hybridised Co-3$d$ and N-2$p$ electrons dominate the DOS around the $E_F$ and thus determine their physical properties. The less-extended Co 3$d$ orbitals yield a narrow band with a high DOS. This probably occurs in SnNCo$_3$ in an enhanced manner. In such a situation, magnetic instability associated with the high DOS is expected[41]. Even a small perturbation such as a physical or chemical pressure or chemical doping might significantly change the properties of SnNCo$_3$, which should be investigated in following studies.

## 4 Reaction mechanism in the melamine method

The reaction conditions in the melamine method in this study are compared to those for the syntheses of binary nitrides and carbides in the previous studies[31,36]. The reaction temperatures for the binary systems are often higher than that in this study. For example, the binary 3$d$-transition-metal nitride VN is synthesised at 1000 °C from V$_2$O$_5$ and melamine at a ratio of V$_2$O$_5$:C$_3$H$_6$N$_6$ = 1:5[31]. A higher reaction temperature of 1200 °C and higher melamine content (V$_2$O$_5$:C$_3$H$_6$N$_6$ = 1:8) are needed for the synthesis of VC[36]. The reaction for the 4$d$ nitride NbN is performed with a nominal composition of Nb$_2$O$_5$:C$_3$H$_6$N$_6$ = 1:3 and reaction temperature of 750 °C, while that for the carbide counterpart NbC is performed at a Nb$_2$O$_5$:C$_3$H$_6$N$_6$ ratio of 1:3 and 1100 °C[31,36]. In contrast, the ternary nitrides and carbides in this study are obtained at approximately 600 and 1000 °C, respectively. On the other hand, the ammonolysis for ZnNNi$_3$ is performed at 600 °C[2]. These characteristics indicate that the anti-perovskite-type compounds are thermodynamically stable and can be synthesised at relatively low temperatures with the aid of an appropriate nitrogen source. It is worth noting that a reaction between Co$_2$O$_3$ or Ni$_2$O$_3$ and melamine in the absence of Zn or Sn does not form a binary nitride or carbide but reduces the oxides to elemental metals[36]. The formation of the thermodynamically stable ternary compound may promote the nitridation or carbidation reactions, which are not possible for the binary systems.

In the case of binary nitrides and carbides, carbides are typically obtained at higher nominal melamine contents and higher temperatures[31,36]. A similar trend is observed in the case of ternary systems. ZnCNi$_3$ and SnCCo$_3$ are synthesised at higher melamine contents and reaction temperatures than those of the corresponding nitrides. We speculate that at high temperatures the formation of a stable N$_2$ molecule reduces the reactive N species and thus terminates the nitridation so that carbidation alternatively occurs.

The nitrogen-source melamine polymerises releasing NH$_3$ gas; intermediate phases such as melam (C$_6$H$_9$N$_{11}$), melem (C$_6$H$_6$N$_{10}$), and melon (C$_{18}$H$_9$N$_{27}$) are produced. Finally, graphitic carbon nitride (g-C$_3$N$_4$) is formed below 520 °C. Above 600 °C, g-C$_3$N$_4$ decomposes to reactive radical species such as C$_3$N$_3^+$, C$_2$N$_2^+$, and C$_3$N$_2^+$[33–35]. As many chemical species are involved in the reaction process, the mechanism of nitridation and carbidation by melamine is not entirely understood. It is speculated that three reaction steps exist: reduction of oxide to metal, nitridation of the metal, and carbidation of the nitride. The reduction of oxides must occur at the first step as several transition-metal oxides are not transformed to nitrides but only reduced to the elemental metals by the melamine reaction. The other two steps have been confirmed by preparations of carbides from reactions between elemental metals and melamine and between nitride and melamine, respectivly[36]. This study shows that the carbides can be synthesised by the reaction between nitrides and melamine at high temperatures. Therefore, it seems that a similar reaction process is involved in the production of ternary compounds. For nitridation, it is suggested that the $NH_3$ gas released from melamine is important in the reaction process. For the binary nitrides, another synthetic route through a reaction between oxides and g-$C_3N_4$ has been investigated[42]. A comparison between the melamine reaction and g-$C_3N_4$ reaction would reveal the role of the $NH_3$ gas in the reaction mechanism.

## IV. CONCLUSION

In summary, we employed nitridation and carbidation reactions using melamine in the syntheses of ternary carbides and nitrides with the anti-perovskite-type structures. By tuning the synthetic conditions such as nominal melamine content and reaction temperature, $ZnNNi_3$ and $ZnCNi_3$ were selectively prepared. Moreover, this method enabled to synthesise a new compound, $SnNCo_3$. The measurements of the physical properties showed that $SnNCo_3$ is a correlated metal with a high DOS at $E_F$. Our findings demonstrate that the melamine method is useful in the synthesis of complex carbides and nitrides and provides a new route to study novel functional materials.

## ACKNOWLEDGEMENTS

This work was partly supported by Japan Society for the Promotion of Science (JSPS) KAKENHI Grant Number JP15K17695, Yazaki Memorial Foundation for Science and Technology, and by Core-to-Core Program (A) Advanced Research Networks.